\documentclass{aa}  

\usepackage{graphicx}
\usepackage{txfonts}
\newcommand{\teff}{T_\mathrm{eff}}
\newcommand{\kms}{ km\,s$^{-1}$}
\begin{document} 

   \title{Zeeman-Doppler imaging of active young solar-type stars
 \thanks{Based on observations made with the HARPSpol instrument on
the ESO 3.6 m telescope at La Silla (Chile), under the program ID 091.D-0836.}}

   \author{T. Hackman \inst{1}
          \and
         J. Lehtinen \inst{1}
          \and
         L. Ros\'en \inst{2}
          \and
         O. Kochukhov \inst{2}
          \and
         M.J. K\"apyl\"a \inst{3}       
          }

   \institute{Department of Physics,
              P.O. Box 64, FI-00014 University of Helsinki\\
              \email{Thomas.Hackman@helsinki.fi}
   \and
   Department of Physics and Astronomy, Uppsala University, 
Box 516, 751 20 Uppsala, Sweden
   \and
   ReSoLVE Centre of Excellence, Aalto University, Department of Computer Science, PO Box 15400, FI-00076 Aalto, Finland}

   \date{Received 7 September 2015 / Accepted 29 December 2015}

 
  \abstract
  {By studying young magnetically active late-type stars, i.e. analogues
to the young Sun, we can draw conclusions on the evolution of the
solar dynamo.}
   {We determine the topology of the surface magnetic field and
study the relation between the magnetic field and cool photospheric spots
in three young late-type stars.}
   {High-resolution spectropolarimetry of the targets was obtained with
the HARPSpol instrument mounted at the ESO 3.6 m telescope. The 
signal-to-noise ratios 
of the Stokes $IV$ measurements were boosted by combining the signal
from a large number of spectroscopic absorption lines through the least
squares deconvolution technique. Surface brightness and magnetic field maps
were calculated using the Zeeman-Doppler imaging technique.}
{All three targets show clear signs of magnetic fields and cool
spots. Only one of the targets, V1358 Ori, shows evidence of the 
dominance of non-axisymmetric modes.
In two of the targets, the poloidal field is significantly stronger
than the toroidal one, indicative of an $\alpha^2$-type dynamo,
in which convective turbulence effects dominate over the weak
differential rotation.
In two of the cases there is a slight anti-correlation between the
cool spots and the strength of the radial magnetic field. However, even in
these cases the correlation is much weaker than in the case of sunspots.}
   {The weak correlation between the measured radial magnetic field and cool 
spots may indicate a more complex magnetic field structure in the spots or
spot groups involving mixed magnetic polarities. Comparison with a previously
published magnetic field map shows that on one of the stars, HD 29615,
the underlying magnetic field changed its polarity between 2009 and 2013.
}


   \maketitle
%

\section{Introduction}

Magnetic fields play a key role in the evolution of solar-type stars. They 
determine the angular momentum loss, shape stellar wind, produce high-energy 
electromagnetic and particle radiation, and influence the energy
balance of planetary atmospheres. 

The general notion as postulated by 
\cite{parker55} is that the solar magnetic
field is generated by an $\alpha \Omega$ -dynamo:
The toroidal field is generated by the winding up and simultaneous
amplification of the poloidal field by differential rotation ($\Omega$
effect). The dynamo loop is closed by the collective inductive and
diffusive effects arising from rotationally influenced convective
turbulence, which generate small-scale poloidal fields from the
underlying toroidal field ($\alpha$ effect), and the enhanced turbulent
diffusion causing efficient reconnection to build up a large-scale
poloidal field ($\beta$ effect).
In rapidly rotating late-type stars, the $\Omega$ effect
is suppressed and therefore other types of dynamos, namely $\alpha^2
\Omega$ or $\alpha^2$, are probably present 
\citep[see e.g.][and references therein]{ossen03}.
In this case, the magnetic field is almost solely sustained by the
$\alpha$ effect generating both poloidal and toroidal fields. Typical
$\alpha \Omega$ solutions show dominance of the toroidal component,
whereas the other types of solutions have more equal energies in
both of the components of the surface field.

Young late-type stars
are generally rapidly rotating owing to the large angular momentum remaining
from the contraction phase during early stellar evolution. Because of magnetic
braking the rotation slows down. This means that as a solar-type star evolves,
its dynamo will move from the $\alpha^2$ or $\alpha^2 \Omega$ -regime
to an $\alpha \Omega$ -dynamo. In order to properly 
understand the mechanism and evolution of the dynamos
of solar-type stars it is essential to study stars of different ages.

To investigate young solar-type stars, we initiated a programme called 
``Active Suns'' for the observation a small sample of young solar 
analogues with the HARPSpol 
instrument mounted at the ESO 3.6 m telescope at La Silla (Chile). The 
three stars
in this study, \object{AH Lep}, \object{HD 29615} and 
\object{V1358 Ori},
are located at the more rapidly rotating and thus more 
magnetically active end of our small sample.

Our primary aims were to determine the topology of the surface magnetic field
and to study the relation between magnetic fields and cool spots. For
this purpose we applied the Zeeman-Doppler imaging (ZDI)
technique using Stokes $IV$ spectropolarimetry with a code developed by
O. Kochukhov \citep{kochukhov14}.

\section{Observations}

Stokes $IV$ spectropolarimetry was collected with the HARPS\-pol 
spectropolarimeter mounted at the ESO 3.6 m telescope at La Silla, Chile 
through the programme ``Active Suns'' (091.D-0836). 
HARPS itself is a fibre-fed, cross-dispersed \'echelle spectrograph described
in greater detail by \cite{mayor03}. The polarisation unit HARPSpol is 
installed
at the Cassegrain focus \citep{piskunov11}. The normal procedure for obtaining
a Stokes $V$ profile is to use four sub-exposures with a rotation of $90
\degr$ of the quarter-wave plate between each sub-exposure. This 
procedure allows the calculation of a ``null spectrum'', which should 
only contain noise and is thus a way to check that there is no spurious 
signal contaminating the Stokes $V$ profile.

The observations were reduced using the REDUCE package \citep{piskunov02}.
The standard reduction  procedure included bias  subtraction, flat fielding  
by a normalised flat field spectrum, subtraction of scattered light, and
an optimal extraction of the spectral orders. 
The blaze function of the
\'echelle  spectrometer was  removed to first order by dividing
the observed stellar spectra by a smoothed spectrum of the
flat field lamp.
The continuum normalisation was performed by fitting a 
second-order polynomial to the blaze-corrected spectra.
A consistent continuum normalisation procedure was applied to the spectra 
extracted from all sub-exposures.
The wavelength solution for each spectrum was accomplished by fitting a 
two-dimensional polynomial, for approximately 1000 extracted thorium
lines observed from the internal lamp assembly.

The spectral resolution of the reduced observations
was $R \approx 105 000$. The rotation phase coverage ($f_\phi$) for each star
was estimated assuming that each observation covers a phase range of 
$[\phi_\mathrm{rot} - 0.05, \phi_\mathrm{rot} + 0.05] $, as suggested by 
\cite{kochukhov13}. The observations are summarised in
Table~\ref{obs} and listed in Table~\ref{allobs}.

Since the aim of the Stokes $V$ observations was to map the surface 
magnetic field, the high demand for the signal-to-noise ratio 
($S/N$) was far beyond what could be reached for individual photospheric 
absorption lines within reasonable exposure times.
In order
to boost the $S/N$ a large number of photospheric
absorption lines in the wavelength range 3800 - 6900 {\AA} were combined with 
least squares deconvolution \citep[LSD;][]{donati97,kochukhov10}. 
The line mask for the LSD procedure was obtained from the VALD data base 
\citep{piskunov95,kupka99}. Regions dominated by strong lines were avoided, 
since the profile of strong lines deviates from the average
line profile. More details on the LSD procedure can be found
in the paper by \cite{kochukhov14}.
The final $S/N$ of the LSD profiles was 5000 -- 12000 in the 
Stokes $V$ spectra. The LSD profiles were calculated on a radial velocity 
grid with a 1.6 km\,s$^{-1}$ step size. The number of spectral lines
used in the LSD mask for each star is listed in Table \ref{obs}. The same
masks were used for both the Stokes $I$ and $V$.

   \begin{figure*}
   \centering
   \includegraphics[width=6.0cm,clip]
{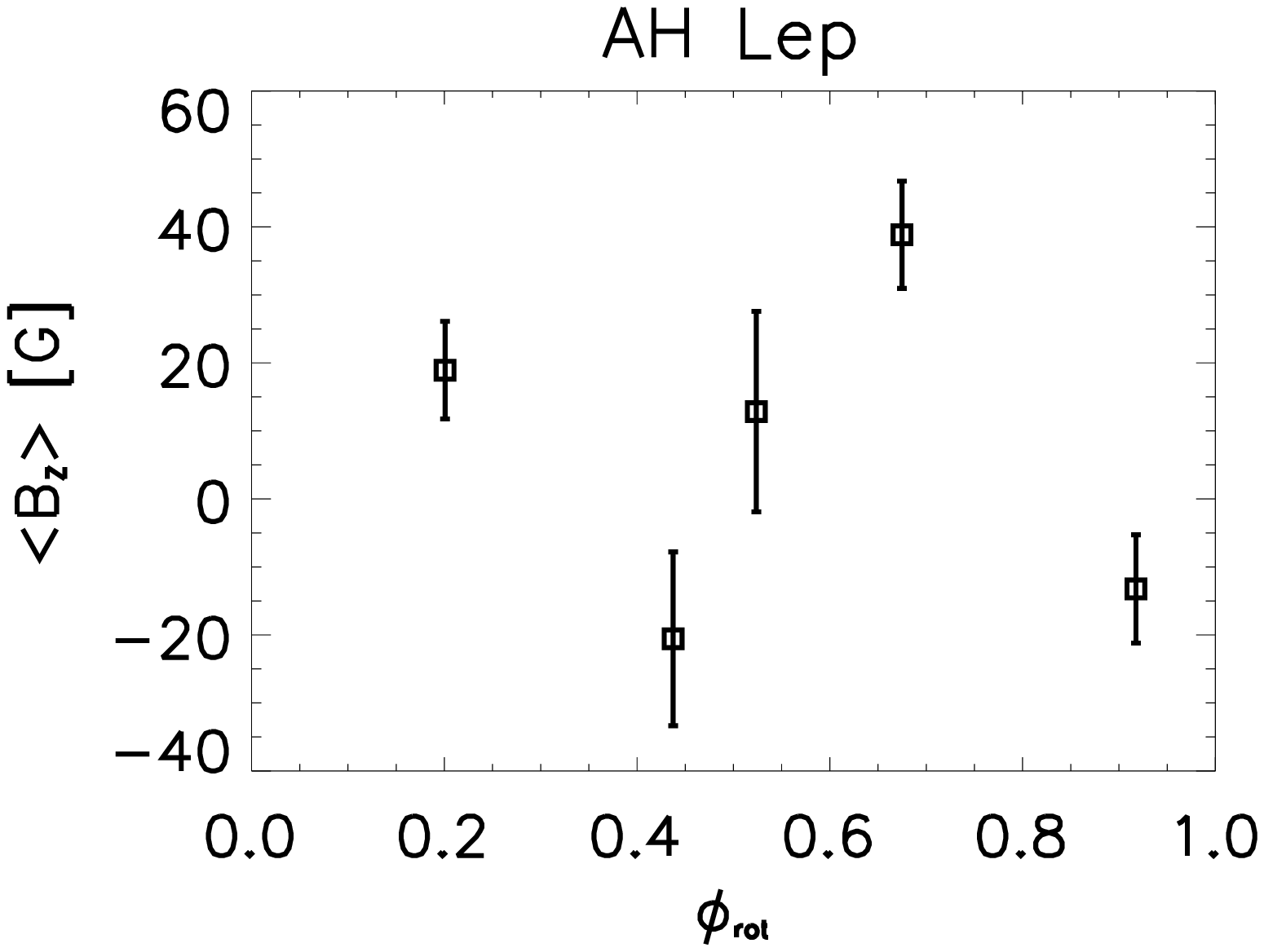}
   \includegraphics[width=6.0cm,clip]
{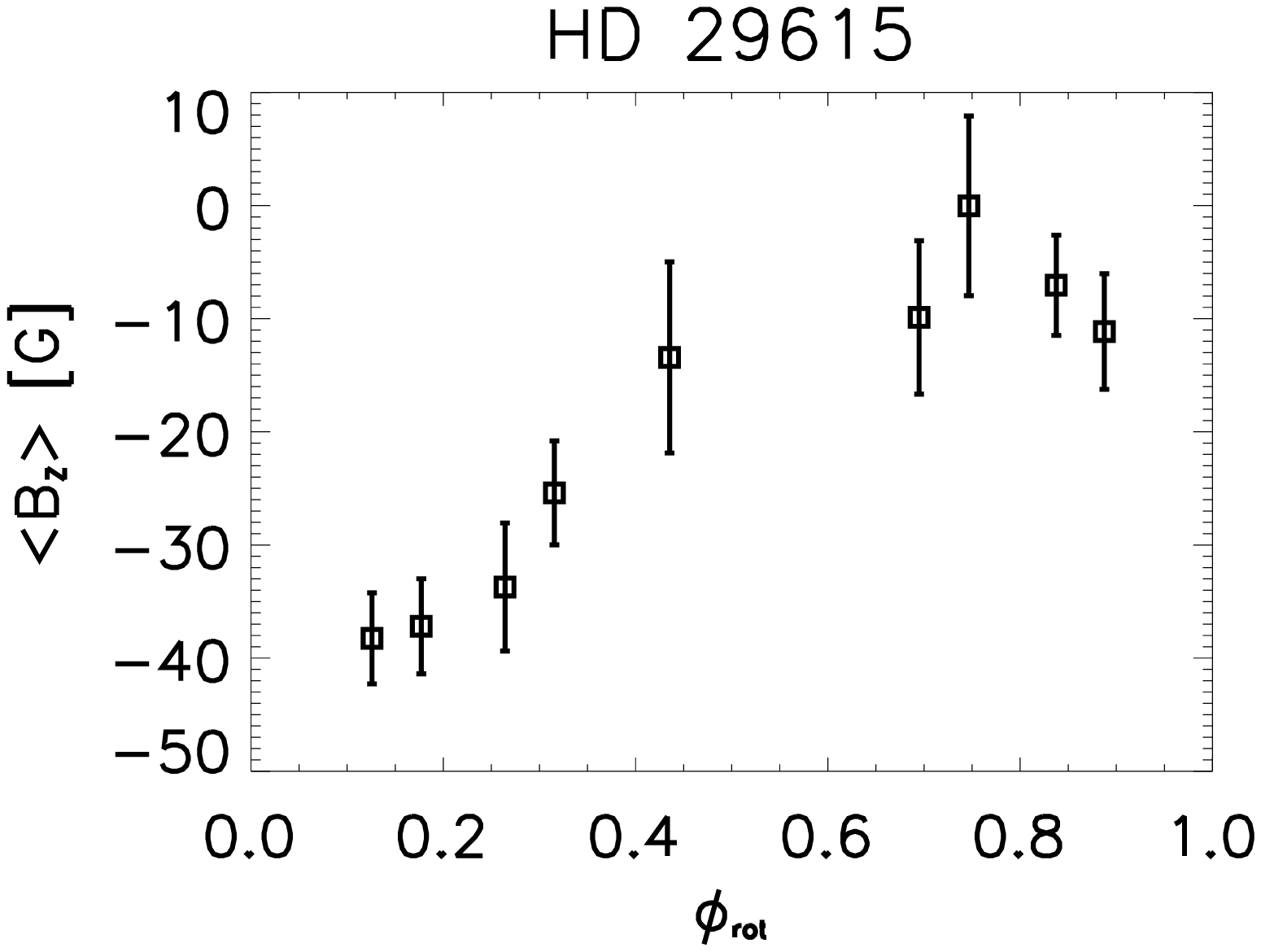}
   \includegraphics[width=6.0cm,clip]
{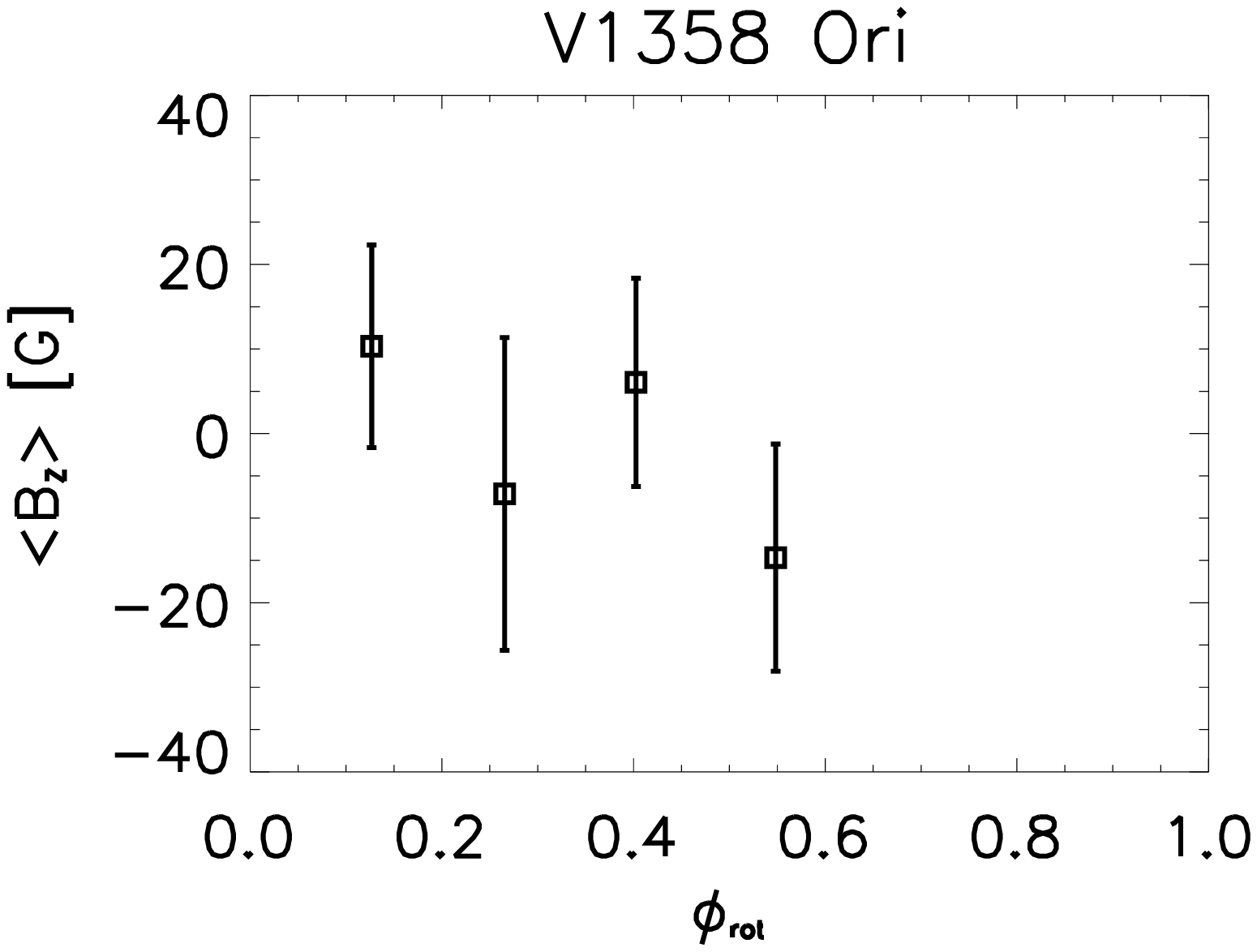}
      \caption{Mean longitudinal magnetic field as a function
of rotational phase. From left to right: AH Lep, HD 29615 and V1358 Ori.}
         \label{bz}
   \end{figure*}

\begin{figure}
\includegraphics[width=2.9cm,clip]
{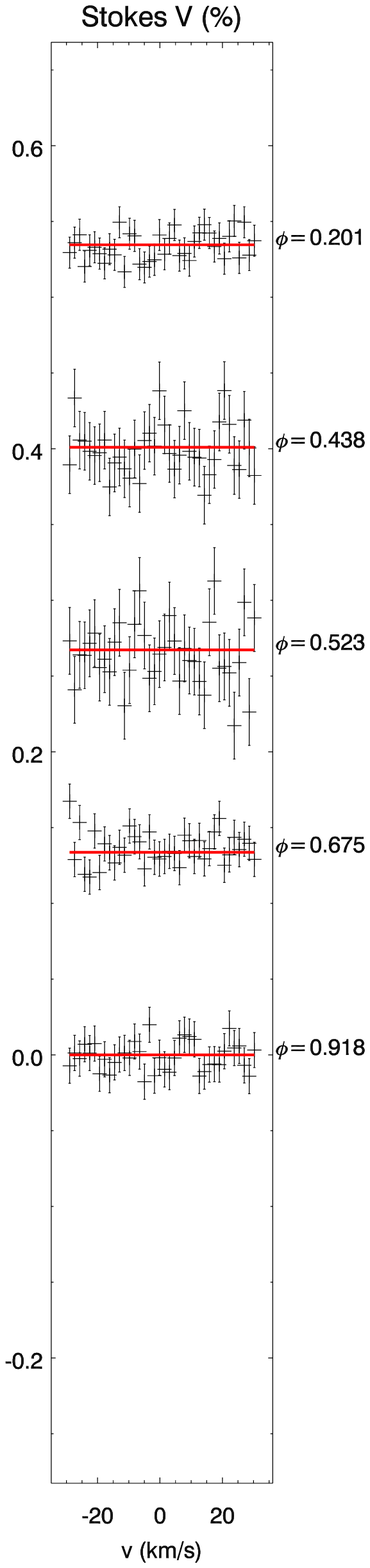}
\includegraphics[width=2.9cm,clip]
{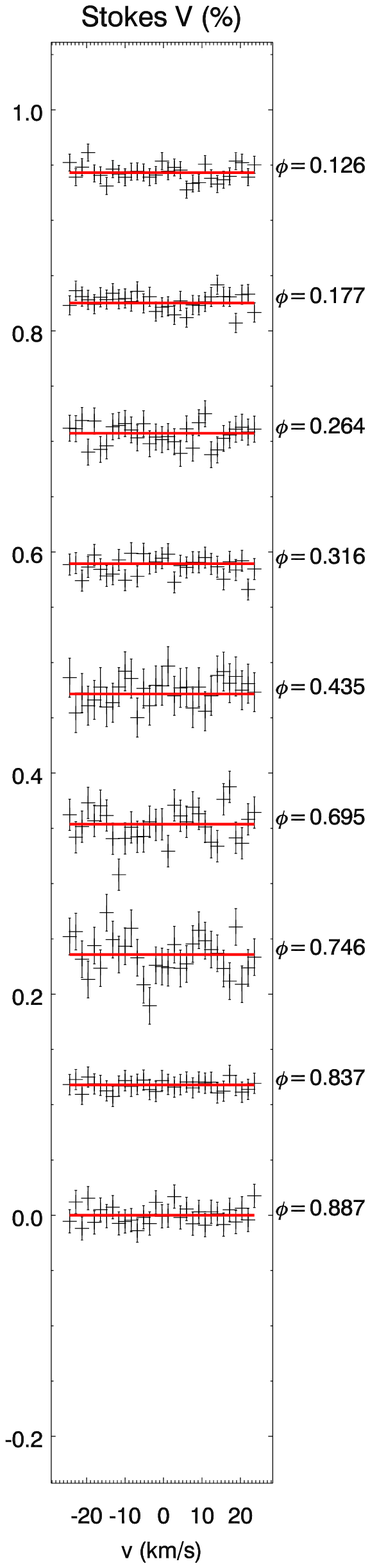}
\includegraphics[width=2.9cm,clip]
{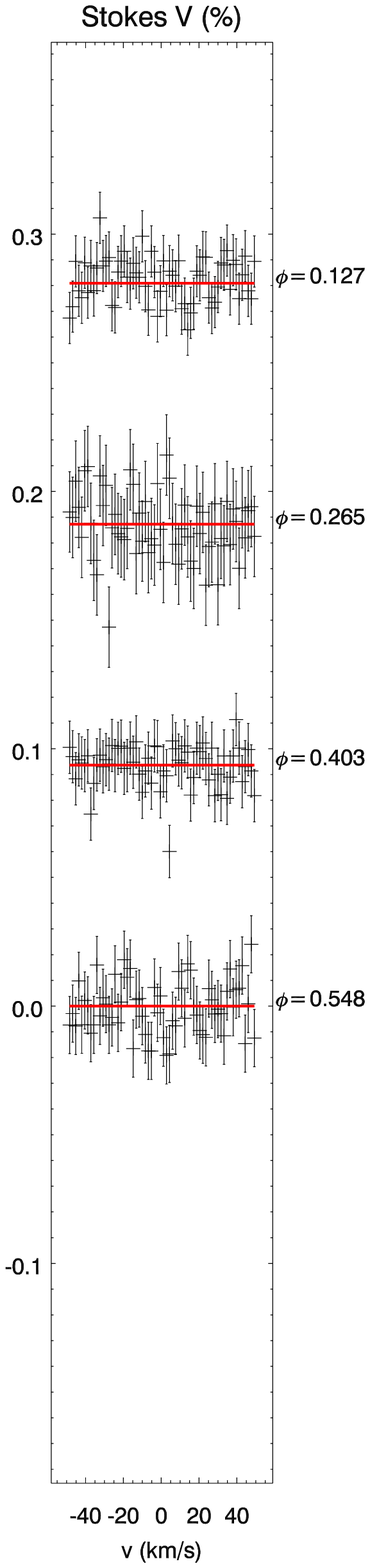}
      \caption{Stokes $V$ null profiles. The error bars are drawn as 
vertical lines through the data points and the 0-level is shown with a red 
line. From left to right: AH Lep, HD 29615, and V1358 Ori.}
         \label{null}
   \end{figure}

From the LSD Stokes $V$ spectra we calculated the false alarm 
probabilities (FAP) for each
observations using the reduced $\chi^2$ statistics 
\citep[see e.g.][]{donati92,rosen13} and the radial velocity range
$[- v \sin i - $7.0\kms, $v \sin i + $7.0\kms$]$. Each point of the Stokes $V$ profile had its own error 
estimate $\sigma_i$ corresponding to the 
$S/N$. This $\sigma_i$ was used for the calculation of the FAP values.
The same procedure was also followed for the null spectra, which naturally reached
the same $S/N$ level as the Stokes $V$.
The minimum FAP values for the Stokes V profiles of the
targets are listed in Table \ref{obs}
and all FAP values for both the Stokes $V$ and the null profiles are 
listed in Table \ref{allobs}.
Using a FAP limit of $10^{-5}$ we 
concluded that all three targets showed definite detection of a magnetic
field. We also evaluated the mean
longitudinal magnetic field $\langle B_Z \rangle$ from the first moment of the 
Stokes $V$ profile as described by \cite{kochukhov10}
for both the Stokes $V$ observations and the null spectra using the same
radial velocity range as for the FAP. These 
measurements are listed in Table \ref{allobs}. The $\langle B_Z 
\rangle$ for each stellar observation as a function of the rotation phase
is plotted in Fig. \ref{bz}. The errors on $\langle B_Z \rangle$ were 
estimated by propagating the errors on the observations.
The null spectra are plotted in Fig. \ref{null}.

\begin{table}
\caption{Summary of observations with HARPSpol at the ESO 3.6 m telescope.}
\centering
\begin{tabular}{lccccr}
\hline \hline
Star & $\langle S/N \rangle$\tablefootmark{a} & $n_\mathrm{LSD}$\tablefootmark{b}
& $n_\phi$ \tablefootmark{c} & $f_\phi$ \tablefootmark{d} & FAP$_{\min}$ 
\tablefootmark{e} \\ 
\hline
AH Lep    & 7400 & 3493 & 5 & 0.49 & $1.4 \cdot 10^{-12}$ \\
HD 29615  & 9100 & 3344 & 9 & 0.68 & $ < 10^{-16}$\\
V1358 Ori & 8800 & 3085 & 4 & 0.40 & $1.7 \cdot 10^{-10}$ \\
\hline
\end{tabular}
\tablefoot{
\tablefoottext{a}{Mean signal-to-noise ratio $\langle S/N \rangle$ of reduced 
Stokes $V$ LSD line profiles, }
\tablefoottext{b}{number of spectral lines in LSD ($n_\mathrm{LSD}$), }
\tablefoottext{c}{number of observed 
phases ($n_\phi$), }
\tablefoottext{d}{estimated phase coverage ($f_\phi$), }
\tablefoottext{e}{minimum FAP for Stokes $V$ signals.}
}

\label{obs}
\end{table}

\begin{table*}
\caption{Observations with HARPSpol at the ESO 3.6 m telescope.}
\centering
\begin{tabular}{llclrrlccc}
\hline \hline
Star & Date & $t$\tablefootmark{a} & $\phi_\mathrm{rot}$ & $t_\mathrm{exp}$ [s] & 
$S/N$ & FAP [$V$] & FAP [null $V$] & $\langle B_Z \rangle$ 
[G] & Null $\langle B_Z \rangle$ [G]\\ 
\hline
AH~Lep & 2013 Sep 9 & 45.82 & 0.918 & 4$\times$700 & 8613 & $1.4\cdot10^{-12}$ &
 0.97 & $-13.2 \pm 8.0$  & $-6.4 \pm 8.0$\\
& 2013 Sep 10 & 46.81 & 0.675 & 4$\times$700 & 8822 & $6.2\cdot10^{-6}$ & 0.72 & $ 38.9 \pm 7.9$ & $-7.0 \pm 7.9$  \\
& 2013 Sep 11 & 47.81 & 0.438 & 4$\times$700 & 5265 & $1.9\cdot10^{-3}$ & 0.94 &
$-20.6 \pm 12.7$ & $0.6 \pm 12.8$\\
& 2013 Sep 12 & 48.81 & 0.201 & 4$\times$700 & 9744 & $1.4\cdot10^{-3}$ & 0.90 &
$18.9 \pm 7.2$ & $-8.6 \pm 7.2$ \\
& 2013 Sep 15 & 51.86 & 0.523 & 4$\times$800 & 4539 & 0.44 & 0.83 & $12.8 
\pm 14.7$ & $13.3 \pm 14.7$ \\
\hline
HD~29615 & 2013 Sep 9 & 45.78 & 0.837 & 4$\times$900 & 10906 & $< 10^{-16}$ &
1.00 & $-7.0 \pm 4.4 $ &  $1.5 \pm 4.4$ \\
& 2013 Sep 9 & 45.90 & 0.887 & 4$\times$900 & 9506 &$< 10^{-16}$& 0.81
& $-11.1 \pm 5.1$ & $-3.2 \pm 5.1$  \\
& 2013 Sep 10 & 46.77 & 0.264 & 4$\times$1000 & 8524 &$< 10^{-16}$&0.86
& $-33.7 \pm 5.7$ & $1.1 \pm 5.6$ \\
& 2013 Sep 10 & 46.89 & 0.316 & 4$\times$1000 & 10523 &$< 10^{-16}$&0.91
& $-25.4 \pm 4.6$ & $1.5 \pm 4.6$ \\
& 2013 Sep 11 & 47.77 & 0.695 & 4$\times$1000 & 7051 & 0.32 & 0.35 
& $-9.9 \pm 6.8$ & $-4.3 \pm 6.8$\\
& 2013 Sep 11 & 47.89 & 0.746 & 4$\times$1000 & 6025 & $5.0\cdot10^{-5}$ & 0.26
& $-0.0 \pm 7.9$ & $9.8 \pm 7.8$\\
& 2013 Sep 12 & 48.77 & 0.126 & 4$\times$1000 & 12081 &$< 10^{-16}$& 0.84
& $-38.3 \pm 4.0$ & $ 2.1 \pm 4.0$ \\\
& 2013 Sep 12 & 48.89 & 0.177 & 4$\times$1000 & 11542 &$< 10^{-16}$ & 0.90
& $-37.2 \pm 4.2$ & $4.1 \pm 4.2$\\
& 2013 Sep 15 & 51.81 & 0.435 & 4$\times$1100 & 5671 &$< 10^{-16}$ & 0.94 &
$-13.4 \pm 8.4$ & $-7.6 \pm 8.4$ \\
\hline
V1358~Ori & 2013 Sep 9 & 45.86 & 0.548 & 4$\times$675 & 8994 & 0.11 & 0.72
& $ -14.7 \pm 13.4$ & -5.4 $\pm 13.4$ \\
& 2013 Sep 10 & 46.85 & 0.403 & 4$\times$675 & 9843 & $1.7\cdot10^{-10}$& 0.98
& $6.1 \pm 12.3$ & $-1.4 \pm 12.3$ \\
& 2013 Sep 11 & 47.85 & 0.265 & 4$\times$675 & 6390 &0.40  & 1.00
& $-7.1 \pm 18.5$ & $21.1 \pm 18.5$ \\
& 2013 Sep 12 & 48.85 & 0.127 & 4$\times$675 & 10050 & $1.5\cdot10^{-2}$ &0.91
& $10.3 \pm 12.0$ & $9.7 \pm 11.9$ \\
\hline
\end{tabular}
\tablefoot{
\tablefoottext{a}{The time of each observation is given as 
$\mathrm{HJD}-2456500$}
}
\label{allobs}
\end{table*}
\setlength{\tabcolsep}{6pt}

\section{Zeeman-Doppler imaging}

The Stokes $V$ profiles were 
modelled assuming the weak field approximation, i.e. that the local Stokes $V$ 
profile is proportional to the effective Land\'e $g$-factor. The Land\'e factor
was determined as suggested by \cite{kochukhov10} and the values are listed
in Table \ref{stars}.

The Stokes $I$ profiles were modelled using a temperature independent 
analytical Voigt function.
For neutral photospheric metal lines, the absorption is usually larger in a
cool spot than in the unspotted photosphere for solar-type stars. This effect 
is, however, of secondary importance because the reduced continuum level in 
the spot will dominate. Thus, for all photospheric absorption lines, the spot 
will
be observable as an ``emission'' bump moving across the line as the star 
rotates. The amplitude of the bump will, of course, depend on the changing
equivalent width of the absorption line. Therefore, the approximation of 
constant
line profile should only lead to a slight error in the retrieved brightness 
of the spots and the overall 
equivalent width of the line (as this usually becomes slightly stronger when 
large cool spots reduce the observed mean $\teff$).

Similar approximations to those used in this study are commonly used in ZDI 
for the 
Stokes $IV$ when applying the LSD technique \citep[see e.g.][]{petit04}. 
An alternative would be to treat the LSD profiles as a 
single line with average parameters or derive the local LSD profile from the 
full spectrum synthesis. The former still involves approximations and so
the advantages are limited. The latter is feasible for early-type stars 
\citep[see e.g.][]{kochukhov14}, but much more demanding
for late-type stars when imaging both the surface magnetic field and 
temperature (or brightness). Furthermore, none of the stars in this study
showed any strong bumps in the Stokes $I$, reducing the error of the
approximation.

We used the inversLSD code developed by O. Kochukhov.
The stellar surface was divided into 1876 spatial elements. The magnetic 
inversion 
was based on a spherical harmonics expansion, as described by
e.g. \cite{kochukhov14} and \cite{rosen15}. One of the advantages with this 
approach is that one can easily determine the amount of energy contained in 
the toroidal and poloidal components of the magnetic field, as well as the 
different $\ell$-modes.

To set the highest allowed $\ell$-mode, different values of $\ell_{\max}$ from 
3 to 10 were tested.
In the solutions with $\ell_{\max} > 5$, smaller scale structures appeared 
even though the fits to the Stokes $V$ observations were not found to be 
significantly improved. This tendency was independent of the 
$v \sin i$-value, which regulates the expected surface resolution for 
temperature Doppler imaging.
This implies that Stokes $V$ alone is not sensitive to 
small-scale features of higher order $\ell$-modes. In fact,
\cite{rosen15} have shown that recovering such details would also require
using linear Stokes $QU$ polarisation measurements. The maximum was 
therefore set to  $\ell_{\max} = 5$ for all stars regardless of their
rotational velocities.

The magnetic field inversion was regularised with a harmonic penalty function, 
while we used Tikhonov regularisation for the brightness inversion
\citep[see e.g.][]{rosen15}.
In addition to the Tikhonov regularisation we also introduced an extra penalty
function to restrain the surface brightness within values of $[0.5, 1.5]$, with
1.0 corresponding to the unspotted surface:

$$
{R_P(b) = \left\{ \begin{array}{ll}
\sum_i c (b_i - b_{\min})^2 &\mbox{ if $b_i<b_{\min}$} \\
0                &\mbox{ if $b_{\min} \le b_i \le b_{\max}$} \\
\sum_i c (b_{\max} - b_i)^2 &\mbox{ if $b_i > b_{\max}$.}
       \end{array} \right.}
$$

\noindent
Here $b_{\min}$ and $b_{\max}$ are the allowed minimum and maximum values for 
the  brightness $b_i$, and $c$ is a constant.
This extra restriction was necessary because using
the approximation of a constant local Stokes $I$ profile made the brightness
inversion insensitive to the mean brightness. 

The inverse solution was obtained using the Levenberg-Marquardt 
minimisation. Here the first and second derivatives of the extra penalty 
function $R_P$ were calculated analytically.
The magnetic energy contained in the axisymmetric and 
non-axisymmetric harmonic components was calculated using the same definition
as in \cite{fares09}: Modes with $m < \ell/2$ were defined as axisymmetric and
those with $m \ge \ell/2$ considered non-axisymmetric.

\section{Observed stars and their adopted stellar parameters}

The three stars in this study are nearby young solar-type and they are 
effectively 
single stars, in the sense that they lack any companions close enough
for detectable interaction. AH Lep and V1358 Ori are
believed to belong to the local association,
implying ages of 20 -- 150 Myr \citep[see e.g.][]{montes01}, while
HD 29615 has been proposed as a member of the Tucana/Horologium association 
indicating an age of $\sim$30 Myr \citep[see e.g.][]{zuckerman11}.
They are, however, slightly hotter than the young Sun.
For one of the stars, V1358 Ori, there are no 
previously published ZDI studies. Previous ZDI maps of AH Lep and HD29615 
were presented by \cite{carter15} and \cite{waite15}.

\subsection{AH Lep}

\object{AH Lep} (\object{HD 36869}, \object{TYC 5916-792-1}) is classified
as a G3V star \citep{montes01} with a TYCHO catalogue parallax of 28.6 mas 
\citep{wichmann03}, i.e. a distance of $r \approx 35 \mathrm{pc}$. \cite{zuckerman11}
reported an effective temperature of $\teff \approx 5800$ K and a radius of 
$R \approx 1.21 R_\odot$. \cite{cutispoto99} calculated a photometric rotation 
period of $P_\mathrm{phot} \approx 1\fd31$. \citet{messina03} reported a 
maximum photometric amplitude of 0.09 in the V magnitude and
an X-ray flux of $\lg (L_\mathrm{X} / L_\mathrm{bol}) = -3.480$. 
Our best fit to the Stokes $I$ 
profile was achieved with a rotation velocity of $v \sin i = 26.3$ {\kms}, i.e. 
close to the value of 27.3 {\kms} measured by \cite{lopez10}. Combining the 
previously measured radius, rotation period and our rotation velocity estimate
we retrieved an inclination of the rotation axis of $i \approx 36 \degr$.
To calculate the rotation phases ($\phi_\mathrm{rot}$ in Tab. \ref{allobs})
of the spectropolarimetric observations, we used the ephemeris

$$ \mathrm{HJD}_{\phi=0} = 2449730.0 + 1.31 \times E.$$

\subsection{HD 29615}

\object{HD 29615} (\object{HIP 21632}, \object{TYC 6467-702-1}) has a 
Simbad spectral classification of G3V \citep{torres06}. Its
Hipparcos parallax is 17.8 mas \citep{leeuwen07} giving a distance of $r 
\approx 56 \mathrm{pc}$. According to \cite{mcdonald12} its effective temperature is
$\teff \approx 5866 K$.
\cite{vidotto14} reported a radius of  $R \approx 0.96 R_\odot$
and a rotation period of $2\fd32$. \citet{messina10} reported a photometric
amplitude of $A_V = 0.08$. We estimated the rotation velocity
to be $v \sin i \approx 18.5$ {\kms} giving an inclination angle of $i \approx
62 \degr$.
Our observations were phased with the ephemeris

$$ \mathrm{HJD}_{\phi=0} = 2449730.0 + 2.32 \times E.$$

Recently \citet{waite15} published a ZDI study of HD 29615, where they also
estimated the differential rotation of the star. Expressed as the
differential rotation coefficient $k=\Delta \Omega / \Omega$, they searched
for the value of $k$ that gave the best fit to the observations and
reported
$k \approx 0.18$ from the magnetic field structures and $k \approx 0.026$ from
the temperature spot structures.
In principle, a discrepancy between these values
could be caused by different anchor depths. However, such a huge difference
seems unlikely. 
Furthermore, theoretical considerations \citep[see e.g.][]
{kr80} and numerical simulations \citep[e.g.][]{kitch11,cole14}
indicate that the magnetic fields of rapidly rotating solar-type stars 
are only weakly influenced by differential rotation, 
and they are therefore expected to have very small values of $k$. 
Thus, and especially since we simultaneously retrieve
both the magnetic field and surface brightness, we did not apply any surface 
differential rotation in this study.

\begin{table}
\caption{Summary of the adopted parameters.}
\centering
\begin{tabular}{lccccc}
\hline \hline
Star & $P_\mathrm{rot}$ & $v \sin i$ & R         & $i$     & Land\'e $g$-factor \\
     & [d]            & [\kms]    & [$R_\odot$] & [$\degr$] &  \\
\hline
AH Lep   & 1.31 &  26.3 & 1.21 & 36 & 1.214 \\
HD 29615 & 2.32 &  18.5 & 0.96 & 62 & 1.214 \\
V1358 Ori & 1.16  & 41.3 & 1.1 & 59  & 1.207 \\

\hline
\end{tabular}
\label{stars}
\end{table}

   \begin{figure*}
   \centering
   \includegraphics[bb=255 40 400 812,width=3.5cm,clip,angle=90]{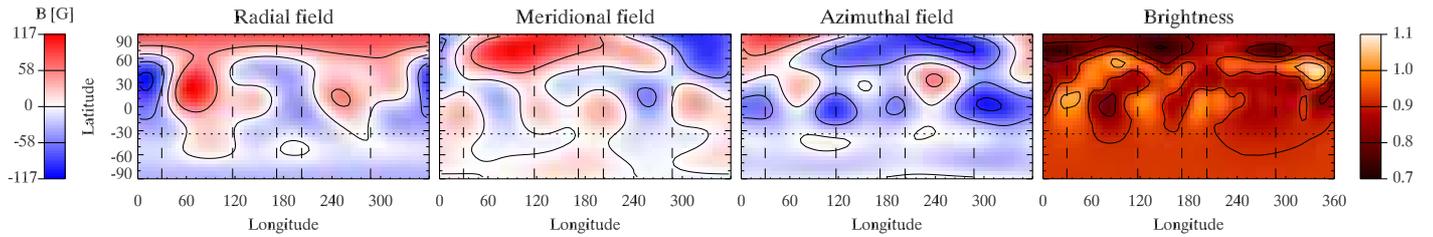}
      \caption{ZDI maps of AH Lep in equirectangular projection. The dashed
vertical lines mark the observed rotation phases and the pointed horizontal
line the limit of visibility due to the stellar inclination.}
         \label{AHLepim}
   \end{figure*}

   \begin{figure*}
   \centering
   \includegraphics[bb=255 40 400 812,width=3.5cm,clip,angle=90]
{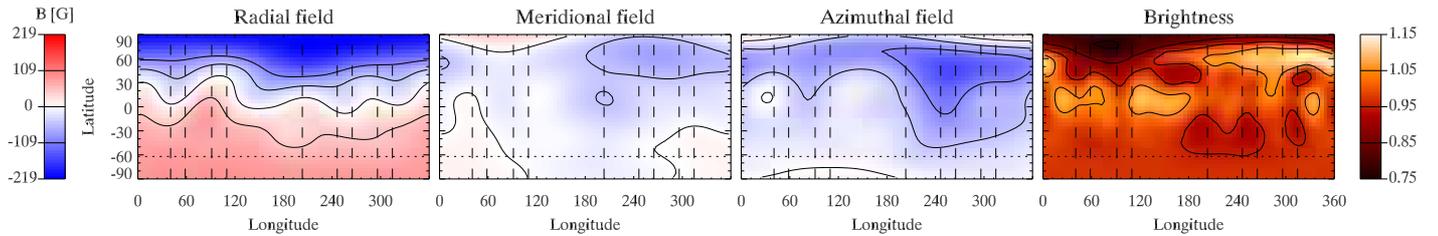}
      \caption{Same as Fig. \ref{AHLepim} for HD 29615}
         \label{HD29615im}
   \end{figure*}

   \begin{figure*}
   \centering
   \includegraphics[bb=255 40 400 812,width=3.5cm,clip,angle=90]
{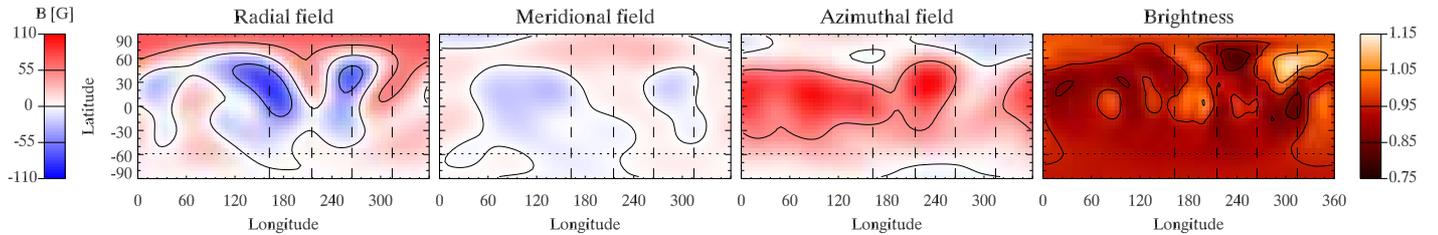}
      \caption{Same as Fig. \ref{AHLepim} for V1358 Ori}
         \label{HD43989im}
   \end{figure*}

   \begin{figure*}
   \centering
   \includegraphics[bb=40 65 355 775,width=5.5cm,clip]
{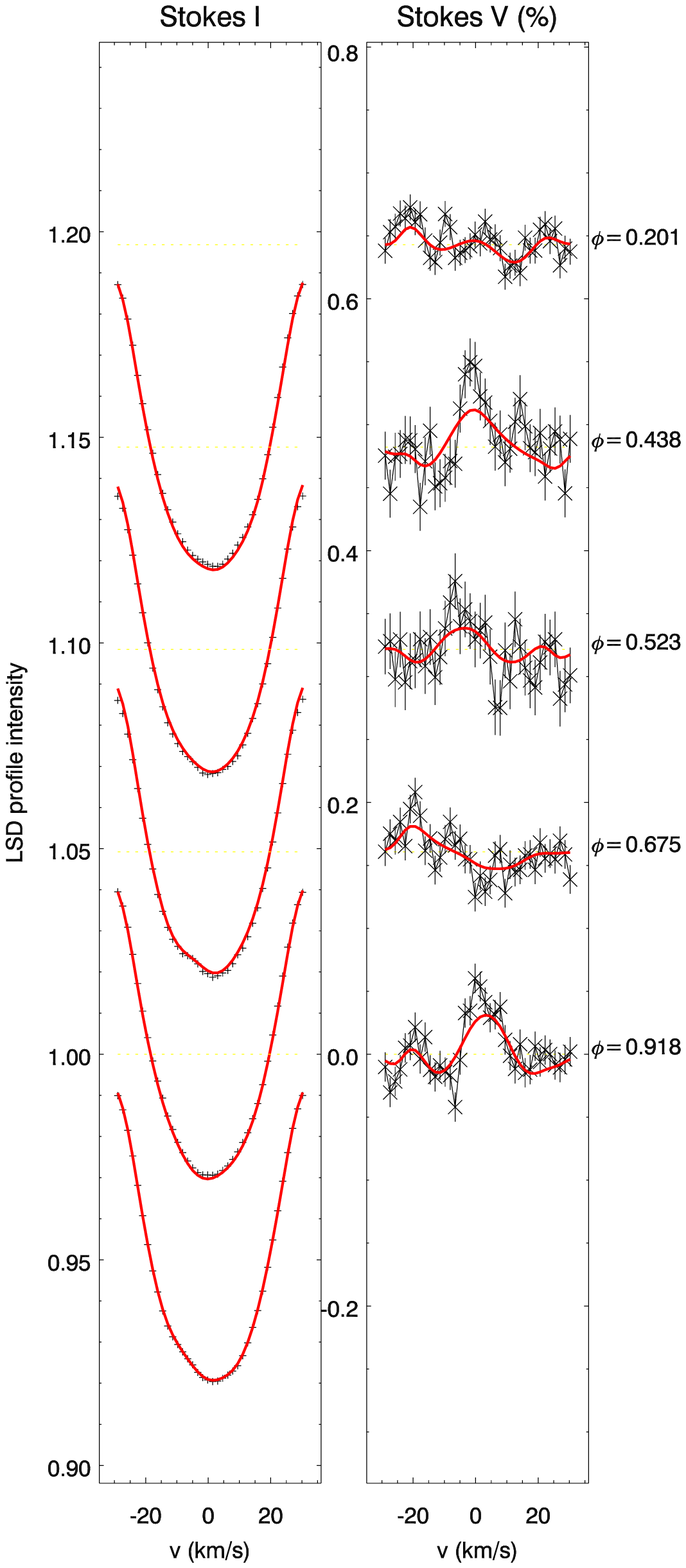}
   \includegraphics[bb=40 65 355 775,width=5.5cm,clip]
{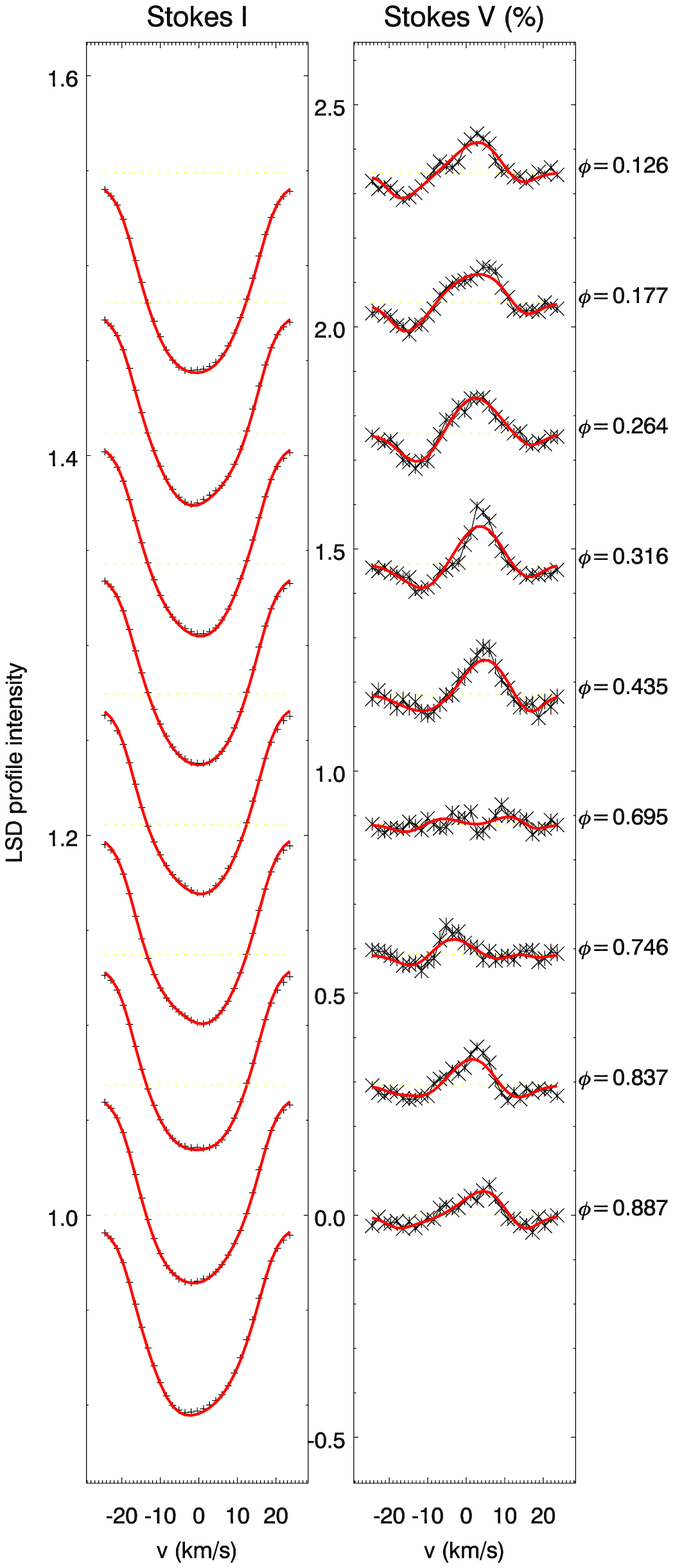}
   \includegraphics[bb=40 65 355 775,width=5.5cm,clip]
{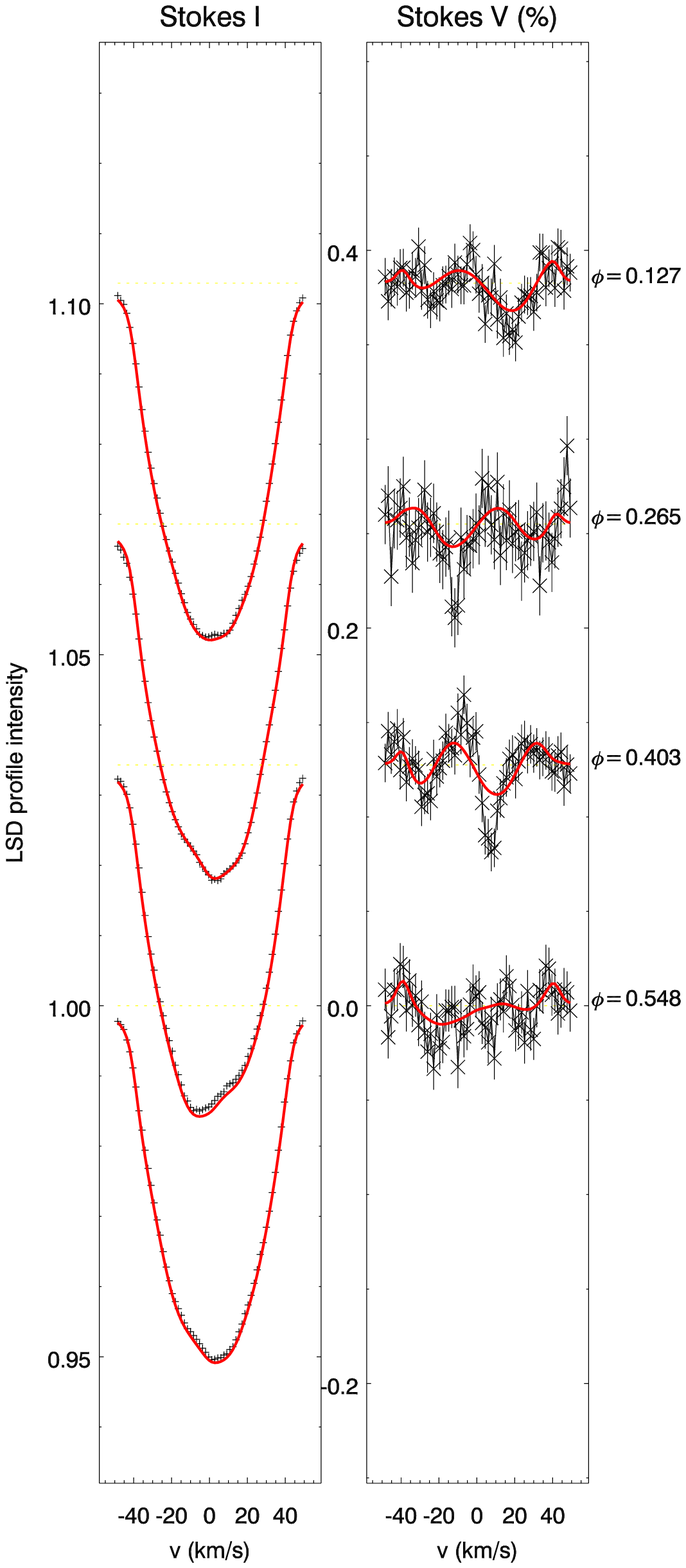}
      \caption{The Stokes $IV$ observations (plus signs/crosses) and 
profiles (solid line in red) calculated from 
the ZDI solutions. The error bars for the Stokes $V$ observations are
drawn as vertical lines through the data points. From left to right: AH Lep, 
HD 29615, and V1358 Ori.}
         \label{prof}
   \end{figure*}

\subsection{V1358 Ori}

\object{V1358 Ori} (\object{HD 43989}, \object{HIP 30030}, 
\object{TYC 4788-1272-1}) was classified as a F9V-type star by \cite{montes01}.
\cite{leeuwen07} reported a Hipparcos parallax of 20.31 mas, placing it at a 
distance of $r \approx 49 \mathrm{pc}$.
\cite{mcdonald12} reported an effective temperature of $\teff \approx 6032$ K, 
while the estimate of \cite{vican14} was $\teff \approx 6100$ K. Furthermore,
\cite{vican14}
reported a radius of $R \approx 1.05 R_{\odot}$, while \cite{zuckerman11} 
presented the value $R \approx 1.08 R_{\odot}$. We adopted the value 1.1 
$R_{\odot}$. From photometric observations \cite{cutispoto03} calculated a 
rotation period of $P_\mathrm{phot} \approx 1\fd16$ and estimated the 
photometric amplitude to be $A_V = 0.08$. Furthermore, they reported an X-ray
flux of  $\lg (L_\mathrm{X} / L_\mathrm{bol}) = -3.629$.
 We estimated the rotation
velocity to be $v \sin i \approx 41.3$ {\kms}, giving an inclination of
$i \approx 59 \degr$. To phase our observations, we used the ephemeris

$$ \mathrm{HJD}_{\phi=0} = 2449681.5 + 1.16 \times E.$$

\section{Results}

The resulting ZDI-maps are presented in Figs. \ref{AHLepim}--\ref{HD43989im} 
and their corresponding Stokes I and V profiles in Fig. \ref{prof}.
The strongest magnetic field is detected on HD 29615 and the weakest field
on V1358 Ori (Table~\ref{magn}). HD 29615 shows an almost axisymmetric 
structure with both the magnetic field and cool spots concentrated at high 
latitudes around the pole. In V1358 Ori the axisymmetric component of the
field is stronger than the non-axisymmetric. 
AH Lep, on the other hand, has clearly non-axisymmetric 
geometries of both the magnetic fields and cool spots. It should be noted, 
however, that the ZDI maps of AH Lep and V1358 Ori are based on poor
phase coverages. 

The poloidal component dominates the surface fields of AH Lep and HD 29615, 
while the toroidal component is stronger for V1358 Ori. In terms of 
complexity, the $\ell = 1$ mode dominates the fields of HD 29615 and 
V1358 Ori, while $\ell = 2$ is strongest for AH Lep (containing 31.2\% of the
magnetic energy). The information on the magnetic field of each 
star is summarised in Table~\ref{magn}. This table lists the maximum detected 
surface field ($B_{\max}$) and the percentage of magnetic energy contained in 
the different components of the field (toroidal/poloidal, 
axisymmetric/non-axisymmetric and  $\ell = 1$ mode).

We also calculated the linear correlation between the magnetic field maps and
surface brightness maps. The correlation
between the radial magnetic field ($\vert B_r \vert$) and surface brightness 
($b$) is especially interesting. In 
the solar case cool spots are, of course, concentrated in regions were strong 
magnetic fields penetrate the surface. This implies a strong negative
correlation $r(\vert B_r \vert,b)$. We could only find a weak correlation for 
the three
stars, and in the case of V1358 Ori the correlation coefficient is
positive, i.e. of an opposite sign than expected
(Table~\ref{magn}). Again, poor phase coverage may play a role.

\begin{table*}
\caption{Summary of magnetic field detection.}
\centering
\begin{tabular}{lrcccccr}
\hline \hline
Star & $B_{\max}$  & Poloidal & Toroidal & Axisymm.&
Non-axisymm. & $\ell = 1$ & $r(\vert B_r \vert,b)$\\
         & [G] & [\%] & [\%] & [\%] & [\%] & [\%] &    \\ 
\hline
AH Lep   & 133   & 58.9 & 41.1 & 38.9 & 61.1 & 27.0 & -0.19 \\
HD 29615 & 245   & 62.8 & 37.2 & 87.6 & 12.4 & 77.2 & -0.26 \\
V1358 Ori & 111  & 31.3 & 68.7 & 72.9 & 27.1 & 70.9 &  0.21 \\
\hline
\end{tabular}

\label{magn}
\end{table*}

\section{Conclusions}

We see considerable differences in the topology of the surface magnetic fields
of the three stars in this study. Since the stars themselves are evolutionarily
and structurally close, it is likely that these differences reflect
temporal changes rather than different dynamo mechanisms. 

We only have three snapshots, which means that
far reaching conclusions cannot be drawn from our study. However, 
the poloidal and toroidal components are of same order of
magnitude in all three targets, which is a strong indication of a dynamo of 
$\alpha^2$ or $\alpha^2 \Omega$ type \citep[cf.][where dynamos of different 
types are compared in direct numerical simulations]{kapyla13}.
On the other hand, the clear dominance of the axisymmetric mode in 
two of the three targets is somewhat unexpected from mean-field dynamo theory 
\citep[e.g.][]{kr80}. Again, this could be due to the non-stationarity over 
time, i.e. caused by the targets being at different phases of the oscillatory 
dynamo cycle. This conclusions is not at odds with the proposed type of the 
dynamo, as even for $\alpha^2$ dynamos, oscillatory solutions can be expected
\citep[see][and references therein]{kapyla13}.

The previous ZDI study of HD 29615 was based on observations collected in 
November --December 2009 \citep{waite15}. These ZDI maps 
look quite similar to our maps. In both cases the magnetic field is dominated
by a strong radial field centred near the rotational pole that covers the 
whole polar region. Interestingly, the  polarity in the 2009 map is opposite 
to that in our map. Thus, the star has undergone
a polarity reversal between December 2009 and September 2013.
The brightness maps similarly show a high latitude spot structure 
clearly offset from the strongest radial field.

When studying the correlation between the magnetic field and temperature spots,
one must remember that the detectability of the field is weighted by the 
surface
brightness. If this were the reason for the dark spots not coinciding with 
strong magnetic fields, one would expect the detected
magnetic field strength to be correlated with the surface brightness.
Only for the case of V1358 Ori do we observe such a weak trend, and even
then this is most likely biased by the poor phase coverage.
On the other hand, \citet{rosen12} showed that the ZDI method
is capable of reproducing both the magnetic field and cool spot structures in 
cases where they coincide. In the case of our maps of AH Lep and especially 
of V1358 Ori,
the poor phase coverage makes the comparison of magnetic and brightness maps
uncertain. However, while some exceptions exist \citep[see e.g.][]{carroll12},
in most cases where surface magnetic field and temperature, brightness, 
or spot 
occupancy maps have been retrieved from the same spectropolarimetric data, there
seems to be a discrepancy between the concentration of cool spots and magnetic
fields \citep[see e.g.][]{jeffers11,kochukhov13,waite15}.

A possible explanation for the lack of correlation between cool spots and a
detected magnetic field could be that the large spots seen in Doppler images
are actually spot groups with mixed polarity. In this case the surface 
resolution of the ZDI is not able to distinguish the detailed magnetic
field configuration. The images would thus only display a larger scale
field.
If this were the case, one would expect the lifetimes of the spot
structures to be very short as many magnetic flux tubes of
opposite polarities packed tightly together would reconnect and decay
rapidly. This would most likely manifest itself through line profile
variations, and the time scale of changes would be helpful in
providing indirect evidence of the topology of the unresolved
structures. Unfortunately, the limited time span of our observations does not
allow for such an analysis.
The situation is further complicated because temperature spots can be generated
without magnetic fields through vortex instability, as demonstrated by e.g. 
\cite{kapyla11} and \cite{mantere11}.

\cite{rosen15} compared ZDI results retrieved using just Stokes $IV$ with the
full set of Stokes $IVQU$ components for the RS CVn star II Peg. They found
few differences in the temperature maps, but significant differences in the
magnetic field solutions. The retrieval of the meridional field component 
especially benefited from using the linear Stokes $QU$ components. Furthermore,
using just Stokes $IV$ resulted in an underestimation of the total magnetic 
energy. Therefore, we can expect that the real surface magnetic fields of our
targets are stronger and much more complex than indicated by this study.

Despite these shortcomings, the results of this study can be used
to retrieve information on possible stellar 
magnetic cycles. Here it is especially interesting to follow the
evolution of the poloidal and toroidal components and the complexity of the 
field, and to detect polarity reversals of the large-scale field. 
Such reversals have
been seen in many stars and with this study we can add HD 29615 to this
group.

\begin{acknowledgements}
This research has made use of the SIMBAD database operated at CDS, Strasbourg,
France. TH was partly supported by the ``Active Suns'' research project at the 
University of Helsinki.
JL was supported by the Vilho, Yrj\"{o}, and Kalle V\"{a}is\"{a}l\"{a} 
Foundation. OK is a Royal Swedish Academy of Sciences Research Fellow, 
supported by grants from the Knut and Alice Wallenberg Foundation and
Swedish Research Council.
MJK acknowledges financial support from the Academy of Finland
Centre of Excellence ReSoLVE (grant No.\ 272157). The authors thank the 
anonymous referee for useful comments that helped improve the manuscript.

\end{acknowledgements}

\bibliographystyle{aa}
\bibliography{hackman_eso}

\end{document}